\documentclass{IEEEtran}
\usepackage{cite,epsfig,amsmath,multirow,url,xcolor,latexsym,amsfonts,amssymb,wasysym,csquotes,cite,array,bm,dblfloatfix,mathtools,flushend}

\begin{document}
\title{Robust AC Transmission Expansion Planning with Load and Renewable Generation Uncertainties}
\author{P. Naga Yasasvi, Abheejeet Mohapatra, and Suresh Chandra Srivastava 
\thanks{The authors are with the Department of Electrical Engineering, Indian Institute of Technology Kanpur, Kanpur 208016, India. (email: yasasvi@iitk.ac.in; abheem@iitk.ac.in; scs@iitk.ac.in)}\vspace{-0.5cm}}

\maketitle
\begin{abstract}
With the integration of Renewable Energy Sources (RESs), the power network must be robust to handle the system's uncertain scenarios. DC Transmission Expansion Planning (TEP) plans are generally not feasible for the AC network. A first bi-level Robust Optimization (RO) solution approach for AC TEP with injection uncertainties in loads and RESs generations is proposed in this paper. It utilizes a well-known convex relaxation and is solved by Benders Decomposition (BD), where the master determines the robust AC TEP plan. A novel dual slave model is developed for the second level of BD, which circumvents the issues in using the Conic Duality Theory (CDT), and aids in the worst-case realization of uncertainties using a novel set of constraints. The developed dual slave is not a mixed-integer problem and is solved using the Primal-Dual Interior Point Method (PDIPM). The proposed work also uses additional linear constraints to reduce BD's computational burden and direct the master towards optimality. Monte-Carlo Simulation (MCS) based verification, via actual nonlinear and non-convex AC Optimal Power Flow (OPF), proves $100\%$ robustness of the TEP plans for all scenarios.
\end{abstract}

\vspace{-0.3cm}
\section*{Nomenclature}
\subsection{TEP}
\begin{IEEEdescription}[\IEEEsetlabelwidth{$P_{gi},Q_{gi}$}]
\item[$\mathbf{N}_b$] Number of buses
\item[$\mathbf{N}_g$] Number of generator buses ($\subset\mathbf{N}_b$)
\item[$\mathbf{\Omega}_s$] Candidate corridors $(i,j)$ with sending bus $i$
\item[$\mathbf{\Omega}_r$] Candidate corridors $(i,j)$ with receiving bus $i$
\item[$\mathbf{\Omega}_s^0$] Base topology lines $(i,j)$ with sending bus $i$
\item[$\mathbf{\Omega}_r^0$] Base topology lines $(i,j)$ with receiving bus $i$
\item[$\mathbf{\Omega}$ ($\mathbf{\Omega}^0$)] $\mathbf{\Omega}_{s}\cup\mathbf{\Omega}_{r}$ ($\mathbf{\Omega}_s^0\cup\mathbf{\Omega}_r^0$)
\item[$\overline{n}_{ij}$] Maximum identical line installations possible between buses $i,j,\forall(i,j)\in\mathbf{\Omega}$
\item[$IC_{ij}$] Cost of candidate line between buses $i,j,\forall(i,j)\in\mathbf{\Omega}$
\item[$x_{ij}^k$] Binary candidate line variable, which is $1$, if $k^{th}$ line is installed between buses $i,j,\forall(i,j)\in\mathbf{\Omega}$, else $0$
\item[$P_{gi},Q_{gi}$] Real and reactive power generation at bus $i$
\item[$C_i(P_{gi})$] Cost of generating $P_{gi}$, with $C_i(P_{gi})=a_iP_{gi}+b_i$
\item[$\Gamma_d$ ($\Gamma_r$)] Curtailment cost of loads (RESs generations)
\item[$P_{di}$ ($P_{ri}$)] Rated real power load (RES generation) at bus $i$
\item[$CP_{di}$] Real power demand curtailment at bus $i$
\item[$CP_{ri}$] RES generation curtailment at bus $i$
\item[$e_i,f_i$] Real and imaginary part of voltage at bus $i$
\item[$\mathbf{N}(i)$] Already and likely to be connected buses to bus $i$
\item[$n_{ij}^0$] Number of lines between buses $i,j,\forall(i,j)\in\mathbf{\Omega}^0$
\item[$b_{i}^{sh}$] Shunt susceptance at bus $i$
\item[$P_{ij}^k$ ($Q_{ij}^k$)] Real (reactive) power flow on $k^{th}$ line to be installed between buses $i,j,\forall(i,j)\in\mathbf{\Omega}$
\item[$\delta_i$] Tangent of the load power factor angle at bus $i$
\item[$M$] Big-M number
\end{IEEEdescription}\vspace{-0.3cm}
\subsection{BD}
\begin{IEEEdescription}[\IEEEusemathlabelsep\IEEEsetlabelwidth{$g,b_{i},b_{i},b^{sh}$}]
\item[$(\mathbf{A},\mathbf{T})$] Matrix coefficients of binary variables in the constraints of robust TEP in compact form 
\item[$(\mathbf{B}_{e},\mathbf{B}_{ie},\mathbf{G})$] Matrix coefficients of continuous variables in the constraints of compact robust TEP
\item[$(\mathbf{F}_m,\mathbf{F}_s,F_c)$] Coefficient vectors of binary, continuous variables and constant in the objective function of robust TEP in compact form, respectively
\item[$(\mathbf{h},\mathbf{t}_e,\mathbf{t}_{ie},\mathbf{r})$] Right hand side constants in the constraints of compact robust TEP 
\item[$\mathbf{H}$] Matrix coefficient in rotated cone constraints of the compact robust TEP
\item[$\mathbf{J}_e, \mathbf{J}_{ie}$] Matrix coefficient of uncertain parameters $\bm{\xi}$ in compact robust TEP
\item[$\mathbf{U}$] Matrix coefficient of the compact robust TEP, that selects the components of continuous variables $\mathbf{y}_s$ equal to $\mathbf{y}_{cone}$ variables in the rotated cone constraint
\item[$\mathbf{y}_m$] Master stage binary variable vector of BD
\item[$(\mathbf{y}_s, \mathbf{y}_{cone})$] Slave stage continuous variables vector of BD
\item[$\bm{\xi}$] Vector of uncertain parameters (where, $\bm{\xi}^d$ represents the $\textit{d}^{th}$ vector and $\xi_d$ represents the $\textit{d}^{th}$ element of the vector)
\item[$\varXi$] Uncertainty set
\item[$SD(\mathbf{y}_m)$] Sum of generation cost and curtailment cost evaluated from the slave problem
\item[{\parbox[t]{0.2\linewidth}{$(\mathbf{z}_{ms},\mathbf{z}_{s\xi},$\\$z_{cij},\bm{\lambda}_{s\xi},\bm{\lambda}_{cs})$}}]  \parbox[t]{1\linewidth}{Vectors of dual variables in the slave sub-problem}\\
\item[$\chi$] Sum of generation cost and curtailment cost evaluated from the master problem
\item[$\bm{u},\bm{\Psi^+},\bm{\Psi^-}$] Auxiliary variable vector in dual problem
\end{IEEEdescription}\normalsize

\section{Introduction}
\IEEEPARstart{E}{xpansion} planning of the transmission network is essential for the reliable and economic operation of power system \cite{gacitua2018comprehensive}. It provides the system planner the cost-effective and timely installation of new lines in the existing network while maintaining a balance between future demand and generation capacities \cite{gu2012coordinating}. For decades, the DC TEP is being used, which generally does not give a feasible solution for the AC networks \cite{puvvada2019efficient}. Also, the integration of RESs, which inherently have uncertain and variable generations, makes evaluation of the AC TEP plans further challenging \cite{gacitua2018comprehensive}. Hence, it is important to develop robust and optimal AC TEP plans in the presence of loads and RESs injection uncertainties. Various short-term and long-term uncertainties in TEP \cite{7491423} are generally handled through Stochastic Programming (SP) \cite{birge2011introduction} and RO \cite{ben2009robust}.

SP optimizes the objective by directly incorporating the uncertainties in the decision process. DC TEP plans are obtained in \cite{freitas2019strategy} while considering multiple generation scenarios. A dynamic scenario clustering technique and BD are used to solve DC TEP in \cite{zhuo2019incorporating} with high RES penetration. However, as reported in \cite{zhuo2019incorporating}, it cannot be used for AC networks without appropriate linearization. The works in \cite{zhan2017fast} and \cite{sun2018objective} use a forward selection algorithm and an objective based clustering, respectively, to reduce the number of scenarios while solving DC TEP. A scenario analysis based approach to solve multi-stage DC TEP is proposed in \cite{maghouli2010scenario} while chance-constrained based approach in \cite{yu2009chance} solves the single-stage DC TEP. \textbf{\emph{Scenario based SP for TEP do not guarantee complete feasibility of the obtained plans for all practical realizations of uncertain parameters \cite{jabr2013optimization}, as infeasibilities are observed when an attempt is made to reduce the investment costs \cite{freitas2019strategy}}}. 

In contrast to SP, RO based TEP requires the uncertain parameters to be represented in intervals \cite{ruiz2015robust}. Bi-level and tri-level based RO approaches for TEP with uncertain loads and generation capacities are proposed in \cite{ruiz2015robust} and \cite{chen2014robust}, respectively. These use the Constraint-and-Column Generation (CCG) strategy, which explicitly uses primal Benders cuts. Another bi-level robust TEP approach based on BD is proposed in \cite{jabr2013robust}. In \cite{minguez2016robust}, the time to solve the RO based TEP with uncertainties in loads and generation capacities is reduced as compared to \cite{ruiz2015robust,chen2014robust} by using the idea that the worst-case costs occur for generation capacities and loads below and above their nominal values, respectively. A new RO approach with linear decision rules is proposed in \cite{dehghan2017adaptive} to meet the requirements of the existing commercial solvers. In \cite{zhang2017robust}, TEP with both long-term and short-term uncertainties is solved by CCG. Further, the work in \cite{dehghan2015robust} presents a CCG approach to solve TEP along with energy storage expansion planning in the presence of both load and wind power uncertainties. An RO and SP based approach to solve a linearized bundled generation and TEP problem in the presence of uncertainties is discussed in \cite{baharvandi2018bundled}. It may be noted that decomposition based strategies, such as BD, to solve RO based TEP can be easily employed when the resulting models are linear \cite{jabr2013robust}. \emph{\textbf{Hence, the RO based TEP approaches in \cite{chen2014robust,ruiz2015robust,jabr2013robust,minguez2016robust,dehghan2017adaptive,zhang2017robust,dehghan2015robust} utilize DC models to obtain robust DC TEP plans that remain feasible for all practical realizations of uncertainties, as defined by the interval sets. DC TEP plans do not guarantee feasible plans for the AC network, which is evident from the results of \cite{puvvada2019efficient}. Thus, AC TEP is favored over DC TEP for the AC network \cite{jabr2013optimization,puvvada2019efficient}. However, to the best of the authors' knowledge (as per the research gap in the literature \cite{RG}), no attempt has been made in the literature for obtaining robust AC TEP plans in the presence of injection uncertainties}}.

\textbf{\emph{Hence, this paper proposes the first bi-level RO model for AC TEP with uncertainties in loads and RESs generations}}. It is solved by BD technique \cite{benders1962partitioning}, where the first level master sub-problem determines the robust AC TEP plans that are feasible for all realizations (even the worst-case) of the uncertainties. In the second level, dual slave sub-problem aids in the worst-case realization of the uncertain loads and RESs generations. However, the worst-case realization of the uncertain parameters is a herculean task in the conventional nonlinear AC TEP \cite{doi}. The well-established convex relaxation model of the primal deterministic AC TEP (without uncertainties) with binary variable representation of candidate lines in \cite{jabr2013optimization} is adopted in this work to circumvent the above-stated issue partially. It is because the convex model in \cite{jabr2013optimization} guarantees a plan for the deterministic AC TEP that satisfies the AC network by use of linear constraints on the bus voltage phase angles. It is to be noted that \emph{convex relaxation of the AC network is not the focus of this paper}. Further, in the second level of BD for worst-case realization of the uncertain injections, \emph{it is observed that the conventional use of CDT \cite{anderson2016duality} to obtain the dual of the convex relaxation based primal (such as in \cite{8338106}) leads to significant Duality Gap (DG) due to weak duality (also discussed in \cite{proof}). The non-zero DG, if ignored, leads to infeasibility in the initial iterations of BD. Thus, \textbf{a novel dual model, which ensures zero DG/ strong duality (derived and proven in \cite{proof}), is proposed}}. The proposed dual is based on the Lagrange function and the associated Karush-Kuhn-Tucker (KKT) conditions of the convex relaxation based primal. It is also observed that certain constraints in the new dual slave model are functions of variables (besides the primal variables), which are also the associated dual variables of the constraints, as against the dual from CDT \cite{anderson2016duality}. To the best of the authors' knowledge, no existing mixed-integer based commercial solver solves problems with such unique constraints. Hence, the proposed dual slave sub-problem for the worst-case realization of uncertain injections does not resort to the commonly used mixed-integer model (such as in \cite{jabr2013robust}). Instead, to aid the master sub-problem in BD for obtaining robust AC TEP plans, a novel set of linear constraints are used in the proposed dual slave sub-problem for the worst-case realization so that the dual slave can be easily solved by PDIPM \cite{torres1998interior}. The proposed work also introduces a means to reduce the computational burden of BD and to direct the master sub-problem towards optimality by introducing additional constraints with the use of Taylor's series concept in the master sub-problem. Thus, the main contribution of the paper is to \textbf{\emph{propose the first bi-level RO based BD approach that gives robust AC TEP plans in the presence of injection (loads and RESs generations) uncertainties}} by using:
\begin{itemize}
\item the proposed novel dual model for the well-established primal convex model of AC TEP in \cite{jabr2013optimization}. It ensures strong duality, as against the dual from CDT \cite{anderson2016duality}.
\item novel linear constraints in the proposed dual slave sub-problem for the worst-case realization of the uncertain injections so that the dual does not resort to the mixed-integer model in \cite{jabr2013robust} and can be easily solved by PDIPM.
\item additional constraints, with the use of Taylor's series, in the master sub-problem to reduce the computational burden of the overall BD based approach.
\end{itemize}

The rest of the paper is as follows. The formulations of deterministic AC TEP, convex relaxation of AC TEP \cite{jabr2013optimization}, and AC TEP with uncertain injections are discussed in Section \ref{sec2}. Section \ref{sec3} presents the proposed bi-level BD based method for obtaining robust AC TEP plans. Numeric results for various test systems with different injection uncertainties are given in Section \ref{sec4}. Section \ref{sec5} concludes the paper.

\section{AC Transmission Expansion Planning (AC TEP)}\label{sec2}
\subsection{Deterministic AC TEP}\label{2.A}
AC TEP, without any uncertainties and with voltages in rectangular coordinates, can be stated as
\begin{align}\label{Dobj}
&\min\sum_{(i,j)\in\mathbf{\Omega}}\sum_{k=1}^{\overline{n}_{ij}}IC_{ij}x_{ij}^k+\sum_{i\in\mathbf{N}_g}C_i(P_{gi})+CR\\\label{Dxij}
&\text{s.t. }x_{ij}^k\in\{0,1\},\forall (i,j)\in\mathbf{\Omega},k=1,.,\overline{n}_{ij}\\\label{Dxijlim}
&x_{ij}^k\leq x_{ij}^{k-1},\forall (i,j)\in\mathbf{\Omega},k=2,.,\overline{n}_{ij}\\\label{Dcii}
&c_{ii}=e_i^2+f_i^2,\forall i\in\mathbf{N}_b\\\label{Dcsij}
&c_{ij}=e_ie_j+f_if_j,s_{ij}=f_ie_j-f_je_i,\forall(i,j)\in\{\mathbf{\Omega}\cup\mathbf{\Omega}^0\}\\\nonumber
&\sum_{j\in\mathbf{N}(i)}\{n_{ij}^0[(c_{ii}-c_{ij})g_{ij}-s_{ij}b_{ij}]+\sum_{k=1}^{\overline{n}_{ij}}P_{ij}^k\}-P_{gi}+\\\label{Dpbal}&P_{di}-CP_{di}
-P_{ri}+CP_{ri}=0,\forall i\in\mathbf{N}_b\\\nonumber
&\sum_{j\in\mathbf{N}(i)}\{n_{ij}^0[(c_{ij}-c_{ii})b_{ij}-c_{ii}b_{ij}^{sh}-s_{ij}g_{ij}]+\sum_{k=1}^{\overline{n}_{ij}}Q_{ij}^k\}-\\\label{Dqbal}
&Q_{gi}+\delta_i(P_{di}-CP_{di})-c_{ii}b_i^{sh}=0,\forall i\in\mathbf{N}_b\\\label{DPij0}
&-n_{ij}^0\overline{P}_{ij}\leq n_{ij}^0[(c_{ii}-c_{ij})g_{ij}\\\nonumber
&\;\;\;\;\;\;\;\;\;\;\;-s_{ij}b_{ij}]\leq n_{ij}^0\overline{P}_{ij},\forall (i,j)\in\mathbf{\Omega}_s^0;<\underline{z}_{pij0},\overline{z}_{pij0}>\\\label{DPji0}
&-n_{ij}^0\overline{P}_{ij}\leq n_{ij}^0[(c_{jj}-c_{ij})g_{ij}\\\nonumber
&\;\;\;\;\;\;\;\;\;\;\;+s_{ij}b_{ij}]\leq n_{ij}^0\overline{P}_{ij},\forall (i,j)\in\mathbf{\Omega}_r^0;<\underline{z}_{pji0},\overline{z}_{pji0}>\\\label{DPijklim}
&-x_{ij}^k\overline{P}_{ij}\leq P_{ij}^k\leq x_{ij}^k\overline{P}_{ij},\forall (i,j)\in\mathbf{\Omega}_s,\\\nonumber
&\;\;\;\;\;\;\;\;\;\;\;\qquad\qquad\qquad\qquad k=1,.,\overline{n}_{ij};<\underline{z}_{pijk},\overline{z}_{pijk}>\\\label{DPjiklim}
&-x_{ij}^k\overline{P}_{ij}\leq P_{ji}^k\leq x_{ij}^k\overline{P}_{ij},\forall (i,j)\in\mathbf{\Omega}_r,\\\nonumber
&\;\;\;\;\;\;\;\;\;\;\;\qquad\qquad\qquad\qquad k=1,.,\overline{n}_{ij};<\underline{z}_{pjik},\overline{z}_{pjik}>\\\label{DQijklim}
&-x_{ij}^kM\leq Q_{ij}^k\leq x_{ij}^kM,\forall (i,j)\in\mathbf{\Omega}_s,\\\nonumber
&\;\;\;\;\;\;\;\;\;\;\;\qquad\qquad\qquad\qquad k=1,.,\overline{n}_{ij};<\underline{z}_{qijk},\overline{z}_{qijk}>\\\label{DQjiklim}
&-x_{ij}^kM\leq Q_{ji}^k\leq x_{ij}^kM,\forall (i,j)\in\mathbf{\Omega}_r,\\\nonumber
&\;\;\;\;\;\;\;\;\;\;\;\qquad\qquad\qquad\qquad k=1,.,\overline{n}_{ij};<\underline{z}_{qjik},\overline{z}_{qjik}>\\\label{DPijk}
&M(x_{ij}^k-1)\leq (c_{ii}-c_{ij})g_{ij}-s_{ij}b_{ij}-P_{ij}^k\\\nonumber
&\leq M(1-x_{ij}^k),\forall (i,j)\in\mathbf{\Omega}_s,k=1,.,\overline{n}_{ij};<\underline{z}^m_{pijk},\overline{z}^m_{pijk}>\\\label{DPjik}
&M(x_{ij}^k-1)\leq (c_{jj}-c_{ij})g_{ij}+s_{ij}b_{ij}-P_{ji}^k\\\nonumber
&\leq M(1-x_{ij}^k),\forall (i,j)\in\mathbf{\Omega}_r,k=1,.,\overline{n}_{ij};<\underline{z}^m_{pjik},\overline{z}^m_{pjik}>\\\label{DQijk}
&M(x_{ij}^k-1)\leq (c_{ii}-c_{ij})b_{ij}+s_{ij}g_{ij}+c_{ii}b_{ij}^{sh}+Q_{ij}^k\\\nonumber
&\leq M(1-x_{ij}^k),\forall (i,j)\in\mathbf{\Omega}_s,k=1,.,\overline{n}_{ij};<\underline{z}^m_{qijk},\overline{z}^m_{qijk}>\\\label{DQjik}
&M(x_{ij}^k-1)\leq (c_{jj}-c_{ij})b_{ij}-s_{ij}g_{ij}+c_{jj}b_{ij}^{sh}+Q_{ji}^k\\\nonumber
&\leq M(1-x_{ij}^k),\forall (i,j)\in\mathbf{\Omega}_r,k=1,.,\overline{n}_{ij};<\underline{z}^m_{qjik},\overline{z}^m_{qjik}>\\\label{DPglim}
&\underline{P}_{gi}\leq P_{gi}\leq \overline{P}_{gi},\forall i=1,.,\mathbf{N}_g;<\underline{z}_{pgi},\overline{z}_{pgi}>\\\label{DQglim}
&\underline{Q}_{gi}\leq Q_{gi}\leq \overline{Q}_{gi},\forall i=1,.,\mathbf{N}_g;<\underline{z}_{qgi},\overline{z}_{qgi}>\\\label{DVlim}
&|\underline{V}_i|^2\leq c_{ii}\leq |\overline{V}_i|^2,\forall i=1,.,\mathbf{N}_b;<\underline{z}_{vi},\overline{z}_{vi}>\\\label{DCPDlim}
&0\leq CP_{di}\leq P_{di},\forall i=1,.,\mathbf{N}_b\\\label{DCPRlim}
&0\leq CP_{ri}\leq P_{ri},\forall i=1,.,\mathbf{N}_b
\end{align}
where $CR=\Gamma_d\sum_{i\in\mathbf{N}_b}CP_{di}+\Gamma_r\sum_{i\in\mathbf{N}_b}CP_{ri}$. For angle reference bus, $f_i=0$. $g_{ij}$, $b_{ij}$, $b_{ij}^{sh}$ denote the conductance, susceptance and half line charging susceptance of line between buses $i$ and $j$, respectively. $\delta_i$ is undefined for buses with no load. \eqref{Dobj} denotes the total cost of expansion, generation and curtailments. \eqref{Dxij} and \eqref{Dxijlim} ensure sequential line installations. \eqref{Dcii} and \eqref{Dcsij} define the auxiliary variables in terms of real and imaginary parts of the bus voltages. \eqref{Dpbal} and \eqref{Dqbal} denote the real and reactive power balance equations at each bus. \eqref{DPij0} and \eqref{DPji0} represent the real power flow limits on each line in the base topology. \eqref{DPijklim} - \eqref{DQjiklim} enforce the real and reactive power flows on the $k^{th}$ line to $0$ if $x_{ij}^k=0$, else they limit their respective flows within specified bounds. Further, \eqref{DPijk} - \eqref{DQjik} enforce the real and reactive power flows on the $k^{th}$ line to be functions of bus voltages and line admittances when $x_{ij}^k=1$. Limits on generations, bus voltage magnitudes, load and RES generation curtailments are given by \eqref{DPglim} - \eqref{DCPRlim}, respectively. Variables within $<>$ in \eqref{DPij0} - \eqref{DVlim} represent the associated dual variables. AC TEP in \eqref{Dobj} - \eqref{DCPRlim} is nonlinear and non-convex due to \eqref{Dcii} and \eqref{Dcsij}. Hence, RO for this formulation is not a straightforward exercise \cite{doi}. For ease in the worst-case realization of the uncertain parameters to obtain robust AC TEP plans, the convex relaxation of AC TEP in \cite{jabr2013optimization} is adopted in this paper, which is discussed next.



\subsection{Convex relaxation of AC TEP \cite{jabr2013optimization}}\label{2.B}
The convex relaxation of deterministic AC TEP in \eqref{Dobj} - \eqref{DCPRlim}, adopted from \cite{jabr2013optimization}, can be stated as
\begin{align}\label{Dobj1}
&\min\;\eqref{Dobj}\;\text{s.t. }\eqref{Dxij}, \eqref{Dxijlim}, \eqref{Dpbal} - \eqref{DCPRlim}\\\label{Dd1ij}
&D_{1ij}=2c_{ij},\forall(i,j)\in\mathbf{\Omega}\cup\mathbf{\Omega}^0;<\alpha_{ij}>\\\label{Dd2ij}
&D_{2ij}=2s_{ij},\forall(i,j)\in\mathbf{\Omega}\cup\mathbf{\Omega}^0;<\beta_{ij}>\\\label{Dd3ij}
&D_{3ij}=c_{ii}-c_{jj},\forall(i,j)\in\mathbf{\Omega}\cup\mathbf{\Omega}^0;<\gamma_{ij}>\\\label{Dd4ij}
&D_{4ij}=c_{ii}+c_{jj},\forall(i,j)\in\mathbf{\Omega}\cup\mathbf{\Omega}^0;<\phi_{ij}>\\\label{Dcone}
&D_{1ij}^2+D_{2ij}^2+D_{3ij}^2\leq D_{4ij}^2,\forall(i,j)\in\mathbf{\Omega}\cup\mathbf{\Omega}^0;\hspace{-0.22em}<\overline{z}_{ij}>\\\label{Dtheta}
&-\pi/2\leq\theta_i\leq\pi/2,\forall i=1,.,\mathbf{N}_b;<\underline{z}_{\theta i},\overline{z}_{\theta i}>\\\label{Dthetaij}
&-n_{ij}^0\epsilon_{\theta}\leq n_{ij}^0(\theta_i-\theta_j-s_{ij})\leq n_{ij}^0\epsilon_{\theta},\forall(i,j)\in\mathbf{\Omega}^0_s;\\\nonumber
&\qquad\qquad\qquad\qquad\qquad\qquad\qquad\qquad
<\underline{z}_{ij0},\overline{z}_{ij0}>\\\label{Dthetaijk}
&-\epsilon_{\theta}-\pi(1-x_{ij}^k)\leq \theta_i-\theta_j-s_{ij}\leq \epsilon_{\theta}\\\nonumber
&\quad+\pi(1-x_{ij}^k),\forall(i,j)\in\mathbf{\Omega}_s,k=1,.,\overline{n}_{ij};<\underline{z}_{ijk},\overline{z}_{ijk}>
 \end{align}
where \eqref{Dcii} and \eqref{Dcsij} are stated in the relaxed form in \eqref{Dd1ij} - \eqref{Dthetaijk}. \eqref{Dd1ij} - \eqref{Dcone} aid in representing an equivalent rotated cone for every existing and possible corridor in the network \cite{8338106}. $\theta_i$ is voltage phase angle at bus $i$. For angle reference bus, $\theta_i=0$. \eqref{Dtheta} - \eqref{Dthetaijk} are additional linear constraints, adopted from \cite{jabr2013optimization}, on bus voltage phase angles and auxiliary variables. It is clearly demonstrated in \cite{jabr2013optimization} that the use of a similar formulation, as in \eqref{Dobj1} - \eqref{Dthetaijk}, guarantees feasible AC TEP plans that satisfy the actual AC network for the deterministic AC TEP. Further, convex relaxation of AC network is not the focus of our work. Hence, in the presence of injection uncertainties, \eqref{Dobj1} - \eqref{Dthetaijk} is utilized to formulate the RO model, in the next section, so that robust AC TEP plans, which satisfy the actual AC network, are obtained. This is also evidenced through the discussion in Section \ref{234324}. As suggested in \cite{jabr2013optimization}, $\epsilon_{\theta}=0.0044rad$ in \eqref{Dthetaij}, \eqref{Dthetaijk} for all test cases in this paper.




\subsection{AC TEP with injection uncertainties}\label{2.C}
An uncertain real power load is stated as $P_{di}+\xi_{di},\forall i\in\mathbf{N}_b$, where $\xi_{di}\in[\underline{\xi}_{di},\overline{\xi}_{di}],\forall i\in\mathbf{N}_b$, defined by the associated interval uncertainty set. Similarly, an uncertain RES generation is stated as $P_{ri}+\xi_{ri},\forall i\in\mathbf{N}_b$, where $\xi_{ri}\in[\underline{\xi}_{ri},\overline{\xi}_{ri}],\forall i\in\mathbf{N}_b$, defined by the associated interval uncertainty set. Thus, AC TEP with uncertain load and RES injections can be stated as
\begin{align}\label{Robj1}
&\min\;\eqref{Dobj}\;\text{s.t. }\eqref{Dxij}, \eqref{Dxijlim}, \eqref{DPij0} - \eqref{DVlim}, \eqref{Dd1ij} - \eqref{Dthetaijk}\\\nonumber
&\sum_{j\in\mathbf{N}(i)}\{n_{ij}^0[(c_{ii}-c_{ij})g_{ij}-s_{ij}b_{ij}]+\sum_{k=1}^{\overline{n}_{ij}}P_{ij}^k\}-P_{gi}+P_{di}+\\\label{Rpbal}
&\xi_{di}-CP_{di}-P_{ri}-\xi_{ri}+CP_{ri}=0,\forall i\in\mathbf{N}_b;<\lambda_{pi}>\\\nonumber
&\hspace{-0.32em}\sum_{j\in\mathbf{N}(i)}\{n_{ij}^0[(c_{ij}-c_{ii})b_{ij}-c_{ii}b_{ij}^{sh}-s_{ij}g_{ij}]+\sum_{k=1}^{\overline{n}_{ij}}Q_{ij}^k\}-Q_{gi}+\\\label{Rqbal}
&\delta_i(P_{di}+\xi_{di}-CP_{di})-c_{ii}b_i^{sh}=0,\forall i\in\mathbf{N}_b;<\lambda_{qi}>\\\label{RCPDlim}
&0\leq CP_{di}\leq P_{di}+\xi_{di},\forall i=1,.,\mathbf{N}_b;<\underline{z}_{pdi},\overline{z}_{pdi}>\\\label{RCPRlim}
&0\leq CP_{ri}\leq P_{ri}+\xi_{ri},\forall i=1,.,\mathbf{N}_b;<\underline{z}_{pri},\overline{z}_{pri}>
\end{align}
where \eqref{Dpbal}, \eqref{Dqbal}, \eqref{DCPDlim}, \eqref{DCPRlim} are respectively rewritten as \eqref{Rpbal} - \eqref{RCPRlim} with associated uncertain parameters. The worst-case scenario of uncertain parameters $\{\xi_{di},\xi_{ri};\forall i=1,.,\mathbf{N}_b\}$, as per their specified interval uncertainty sets, needs to be realized to obtain robust TEP plans \cite{jabr2013robust}. It can be observed that the uncertain parameters appear in the equality and inequality constraints in \eqref{Rpbal} - \eqref{RCPRlim}. The proposed bi-level BD based approach to obtain robust AC TEP plans from RO counterpart of \eqref{Robj1} - \eqref{RCPRlim} is discussed next.



\section{Proposed Solution Methodology}\label{sec3}
A generic BD based approach \cite{benders1962partitioning} is used to solve for robust AC TEP plans with uncertainties in \eqref{Robj1} - \eqref{RCPRlim}. The associated bi-level RO counterpart can be stated, in a compact form, as
\begin{align}\label{Gobj}
&\min\limits_{\mathbf{y}_m,\mathbf{y}_{cone},\mathbf{y}_s}\{\mathbf{F}_m^T\mathbf{y}_m+\max\limits_{\bm{\xi}\in\bm{\varXi}}\;(\mathbf{F}_s^T\mathbf{y}_s+F_c)\}\\\label{Gineqxi}
\text{s.t. }&\mathbf{A}\mathbf{y}_m\leq\mathbf{h}\\\label{Gineqxci}
&\mathbf{T}\mathbf{y}_m+\mathbf{G}\mathbf{y}_s\leq\mathbf{r}\\\label{Geqxc}
&\mathbf{B}_e\mathbf{y}_s+\mathbf{J}_e\bm{\xi}=\mathbf{t}_e,\forall\bm{\xi}\in\bm{\varXi};<\bm{\lambda}_{s\xi}>\\\label{Gineqxc}
&\mathbf{B}_{ie}\mathbf{y}_s+\mathbf{J}_{ie}\bm{\xi}\leq\mathbf{t}_{ie},\forall\bm{\xi}\in\bm{\varXi};<\mathbf{z}_{s\xi}>\\\label{Geqcone}
&\mathbf{y}_{cone}+\mathbf{U}\mathbf{y}_s=0;<\bm{\lambda}_{cs}>\\\nonumber
&\mathbf{y}_{coneij}\in\mathbf{K},\mathbf{K}=\{\mathbf{y}_{coneij}:\mathbf{y}_{coneij}^T\mathbf{H}\mathbf{y}_{coneij}\leq0\},\\\label{Gcone}
&\forall(i,j)\in\mathbf{\Omega}\cup\mathbf{\Omega}^0;<z_{cij}>
\end{align}
where $\mathbf{y}_m=\{x_{ij}^k,\forall(i,j)\in\mathbf{\Omega},k=1,.,n_{ij}\}$ is vector of first level (master) variables. $\mathbf{y}_{cone}=\{\mathbf{y}_{coneij},\forall(i,j)\in\mathbf{\Omega}\cup\mathbf{\Omega}^0\}$ is the vector of second level (slave sub-problem) variables used to represent the rotated cones. Thus, $\mathbf{y}_{coneij}=\{D_{1ij},D_{2ij},D_{3ij},D_{4ij}\}$. $\mathbf{y}_s$ is vector of other second level (slave) continuous variables. $\bm{\xi}=\{\xi_{d1},.,\xi_{d\mathbf{N}_b},\xi_{r1},.,\xi_{r\mathbf{N}_b}\}$ is the uncertainty vector and $\bm{\varXi}=[\underline{\bm{\xi}},\overline{\bm{\xi}}]$, with $\underline{\bm{\xi}}$, $\overline{\bm{\xi}}$ being the respective lower and upper bounds of the interval uncertainty sets. The objective in \eqref{Gobj} is maximized with respect to $\bm{\xi}$ so that a robust AC TEP plan is obtained for the worst-case realization of uncertain parameters, which is expected to remain feasible for other realizations as well \cite{jabr2013robust}. Thus, with respect to \eqref{Robj1} - \eqref{RCPRlim}, \eqref{Gobj} refers to \eqref{Dobj}. \eqref{Gineqxi} refers to \eqref{Dxij} and \eqref{Dxijlim}. \eqref{Gineqxci} represents \eqref{DPijklim} - \eqref{DQjik} and \eqref{Dthetaijk}. \eqref{Geqxc} corresponds to \eqref{Rpbal} and \eqref{Rqbal}. \eqref{Gineqxc} represents \eqref{DPij0}, \eqref{DPji0}, \eqref{DPglim} - \eqref{DVlim}, \eqref{Dtheta}, \eqref{Dthetaij}, \eqref{RCPDlim} and \eqref{RCPRlim}. \eqref{Geqcone} is equivalent to \eqref{Dd1ij} - \eqref{Dd4ij}. \eqref{Gcone} represents \eqref{Dcone}, where $\mathbf{H}$ is a $4\times4$ diagonal matrix with the first three diagonal elements as $1$ and the last diagonal element is $-1$. Other matrices and vectors in \eqref{Gobj} - \eqref{Geqcone} can be similarly inferred. Vectors within $<>$ are the respective dual variables vectors. \eqref{Gobj} - \eqref{Gcone}, for fixed and feasible $\mathbf{y}_m$, reduce to the following min-max primal slave sub-problem
\begin{align}\label{Pobj}
&\min\limits_{\mathbf{y}_{cone},\mathbf{y}_s}\max\limits_{\bm{\xi}\in\bm{\varXi}}\;(\mathbf{F}_s^T\mathbf{y}_s+F_c)\;\text{s.t. }\eqref{Geqxc} - \eqref{Gcone}\\\label{Pineqxci}
&\mathbf{G}\mathbf{y}_s\leq\mathbf{r}-\mathbf{T}\mathbf{y}_m;<\mathbf{z}_{ms}>
\end{align}
which is a convex programming problem with respect to $\mathbf{y}_{cone}$ and $\mathbf{y}_s$ \cite{boyd2004convex}. RO strongly emphasizes that for a linear and convex model, each uncertain parameter is either at the respective upper or lower bound for the worst-case realization \cite{wu2008interval}. An alternate approach for the worst-case realization of $\bm{\xi}$ is to solve an equivalent dual slave sub-problem of \eqref{Pobj}, \eqref{Pineqxci}, which is discussed next.
\vspace{-0.3cm}

\subsection{Proposed Dual slave sub-problem}
The dual slave sub-problem offers two advantages over the associated primal slave sub-problem: 1) realization of the worst-case scenario of uncertainties is easy, as the uncertain vector $\bm{\xi}\in\bm{\varXi}$ appears only in the objective of the former, 2) search space of the former sub-problem is independent of master sub-problem variables $\mathbf{y}_m$ \cite{jabr2013robust}. Further, for a linear problem, CDT \cite{anderson2016duality} gives an equivalent dual of the associated primal problem with the guarantee of strong duality. The same, however, is not guaranteed for general convex programming problems \cite{anderson2016duality,boyd2004convex} due to weak duality. \emph{This is further discussed in Section 4 of \cite{proof} and also evidenced in the case of AC TEP in Section \ref{sec41}}.

Hence, a novel dual slave sub-problem is proposed for the primal slave sub-problem in \eqref{Pobj}, \eqref{Pineqxci} from the KKT conditions of the associated Lagrange that ensures strong duality/ zero DG. \emph{Associated derivation and proof of strong duality of the proposed dual slave sub-problem are given Sections 2 and 3 of \cite{proof}, respectively}. For convex programming problems, such as in \eqref{Pobj}, \eqref{Pineqxci}, the use of Lagrange dual from associated KKT conditions over the dual from CDT offers the advantage of strong duality. It is also evidenced from the numeric results in Section \ref{sec41}. Thus, the proposed novel dual slave sub-problem for \eqref{Pobj}, \eqref{Pineqxci} is
\begin{equation}\label{Sobj}
\max_{\begin{array}{c}\scriptstyle \mathbf{z}_{ms},\bm{\lambda}_{s\xi},\mathbf{z}_{s\xi},\bm{\lambda}_{cs}, \\
\scriptstyle \mathbf{z}_c,\mathbf{y}_{cone},\bm{\xi}\in \bm{\varXi}\end{array}}\hspace{-0.2cm} SD(\mathbf{y}_m)=\begin{aligned}[t]
&F_c+(\mathbf{T}\mathbf{y}_m-\mathbf{r})^T \mathbf{z}_{ms}-\mathbf{t}_e^T\bm{\lambda}_{s\xi}\\
-&\mathbf{t}_{ie}^T\mathbf{z}_{s\xi}+\bm{\xi}^T(\mathbf{J}_e^T\bm{\lambda}_{s\xi}+\mathbf{J}_{ie}^T\mathbf{z}_{s\xi})
\end{aligned}
\end{equation}
\begin{equation}\label{Seq}
\text{s.t.}\quad \mathbf{G}^T\mathbf{z}_{ms}+\mathbf{B}_e^T\bm{\lambda}_{s\xi}+\mathbf{B}_{ie}^T\mathbf{z}_{s\xi}+\mathbf{U}^T\bm{\lambda}_{cs}+\mathbf{F}_s=0; \langle \mathbf{y}_s\rangle
\end{equation}
\begin{equation}\label{Scone}
\bm{\lambda}_{csij} + 2z_{cij}\mathbf{H}\mathbf{y}_{coneij}=0,\forall(i,j)\in\mathbf{\Omega}\cup\mathbf{\Omega}^0; \langle \mathbf{y}_{coneij}\rangle
\end{equation}
\begin{equation}\label{Slim}
\mathbf{z}_{ms},\mathbf{z}_{s\xi},\mathbf{z}_c\geq0,\;\bm{\lambda}_{s\xi},\bm{\lambda}_{cs},\mathbf{y}_{cone}\text{ unrestricted}
\end{equation}
where $\mathbf{z}_{c}=\{z_{cij},\forall(i,j)\in\mathbf{\Omega}\cup\mathbf{\Omega}^0\}$, $\bm{\lambda}_{cs}=\{\bm{\lambda}_{csij},\forall(i,j)\in\mathbf{\Omega}\cup\mathbf{\Omega}^0\}$ and $\bm{\lambda}_{csij}=\{\alpha_{ij},\beta_{ij},\gamma_{ij},\phi_{ij}\}$ from \eqref{Dd1ij} - \eqref{Dd4ij}. \emph{The associated full dual slave of \eqref{Sobj} - \eqref{Slim} and more details on the derivation are given in \cite{proof}}. The uniqueness of this dual is that the constraints are functions of both primal and dual variables (each constraint in \eqref{Scone} is function of $\mathbf{y}_{coneij}$ that also are the associated dual variables), which, however, is not the case in the dual from CDT \cite{anderson2016duality}. The proposed dual in \eqref{Sobj} - \eqref{Slim} can be easily solved by PDIPM \cite{torres1998interior}.

Consequently, it can be observed from \eqref{Sobj} - \eqref{Slim} that the uncertain vector $\bm{\xi}\in \bm{\varXi}$ appears only in the objective, which is to be maximized. The worst-case uncertain scenario is realized when each element of $\bm{\xi}$\ is either at its maximum or minimum \cite{wu2008interval}. Hence, to attain an optimal value of \eqref{Sobj}, it is evident that a negative (non-negative) element of $\bm{\Psi}=\mathbf{J}_e^T\bm{\lambda}_{s\xi}+\mathbf{J}_{ie}^T\mathbf{z}_{s\xi}$ in \eqref{Sobj} should correspond to the associated element of $\bm{\xi}$ being at its minimum (maximum), else a non-optimal solution is obtained \cite{wu2008interval}. Thus, the associated novel dual slave sub-problem for the worst-case uncertainty realization in \eqref{Sobj} - \eqref{Slim} (second level of BD) is
\begin{equation}\label{MSobj}
\hspace{-0.3cm}\max_{\begin{array}{c}\scriptstyle \bm{\Psi^+}, \bm{\Psi^-},\mathbf{u},\mathbf{z}_{ms},\bm{\lambda}_{s\xi}, \\[-4pt]
\scriptstyle \mathbf{z}_{s\xi},\bm{\lambda}_{cs},\mathbf{z}_c,\mathbf{y}_{cone}\end{array}}\hspace{-0.2cm}\begin{aligned}[t] &SD(\mathbf{y}_m)=F_c+(\mathbf{T}\mathbf{y}_m-\mathbf{r})^T \mathbf{z}_{ms}\\
&-\mathbf{t}_e^T\bm{\lambda}_{s\xi}-\mathbf{t}_{ie}^T\mathbf{z}_{s\xi}+\bm{1}^T\mathbf{u}\;\text{s.t. }\eqref{Seq}\text{ - }\eqref{Slim}\end{aligned}
\end{equation}
\begin{equation}\label{MSpsi}
\Psi_k^+-\Psi_k^--\Psi_k=0,\forall k=1,.,|\bm{\xi}|
\end{equation}
\begin{equation}\label{MSpsilim}
0 \le \Psi_k^+ ,\Psi_k^- \le L,\, \forall k=1,.,|\bm{\xi}|
\end{equation}
\begin{equation}\label{MSuzeta}
-\Psi_k^-\xi^{max}_k+\Psi_k^+\xi^{min}_k \le u_k \le \Psi_k^+\xi^{max}_k -\Psi_k^-\xi^{min}_k,\,\\
\forall k=1,.,|\bm{\xi}|
\end{equation}
where $\bm{1}$ and $\mathbf{u}$ are vectors of same dimension as $\bm{\xi}$, i.e., $|\bm{\xi}|$. $\Psi_k$ is the $k^{th}$ element of $\bm{\Psi}$. $\Psi_k$ is coefficient of ${\xi}_k$, the $k^{th}$ element of $\bm{\xi}$, in \eqref{Sobj}. $L$ is a Big-M number. Based on the sign of $\Psi_k\forall k=1,.,|\bm{\xi}|$, $\Psi_k^+$ and $\Psi_k^-$ take appropriate values as per \eqref{MSpsi} and \eqref{MSpsilim}. Based on $\Psi_k^+$ and $\Psi_k^-$, $u_k\forall k=1,.,|\bm{\xi}|$ chooses appropriate value, as in \eqref{MSuzeta}, so that the objective is maximized in \eqref{MSobj}. To the best of the authors' knowledge, no existing mixed-integer based commercial solver solves problems with constraints such as in \eqref{Scone}, which are also part of the dual slave sub-problem for the worst-case uncertainty realization. By use of the linear constraints \eqref{MSpsi} - \eqref{MSuzeta} in the proposed dual slave sub-problem for the worst-case uncertainty realization that also has the unique constraints in \eqref{Scone}, PDIPM \cite{torres1998interior} can be easily used for obtaining the associated solution. Typically, $L\approx30M$ works well in this case. After solving \eqref{MSobj} - \eqref{MSuzeta}, if $\Psi_k^+-\Psi_k^-=\Psi_k\ge0$, then $\xi_k=\xi^{max}_k$ and associated $u_k$ is reset as $u_k=\Psi_k\xi^{max}_k$, else $\xi_k=\xi^{min}_k$ and associated $u_k$ is reset as $u_k=\Psi_k\xi^{min}_k$. This is done $\forall k=1,.,|\bm{\xi}|$, and as a result, $SD(\mathbf{y}_m)$ is recalculated with the reset value of $\mathbf{u}$ and converged values of other variables in \eqref{MSobj}. Based on $SD(\mathbf{y}_m)$ and other variables in \eqref{MSobj} - \eqref{MSuzeta}, the master sub-problem is solved to obtain the new master variables $\mathbf{y}_m$ for next iteration of BD. Also, due to the load and RES curtailments in \eqref{Pobj}, \eqref{Pineqxci}, \eqref{MSobj} - \eqref{MSuzeta} is always a bounded problem with the guarantee of a feasible solution \cite{proof,jabr2013robust}.
\vspace{-0.3cm}
\subsection{Master sub-problem}
Let $\mathbf{z}_{ms}^p$, $\bm{\lambda}_{s\xi}^p$, $\mathbf{z}_{s\xi}^p$ and $\mathbf{y}_{cone}^p$ be the converged values of the dual slave sub-problem in \eqref{MSobj} - \eqref{MSuzeta} in the $p^{th}$ iteration of BD. Also, let the associated worst-case $\bm{\xi}$ be $\bm{\xi}^p$. By using these values, an optimality cut is evaluated (as dual slave always gives a feasible solution), and added to the master sub-problem so as to improve $\mathbf{y}_m$ in the next iteration of BD. The compact master sub-problem in the $p^{th}$ iteration of BD is
\begin{align}\label{Mobj}
&\min\limits_{\mathbf{y}_m,\chi,\mathbf{y}_s,\Delta\mathbf{y}_{cone}}\mathbf{F}_m^T\mathbf{y}_m+\chi\text{ s.t. }\eqref{Gineqxi},\eqref{Gineqxci}\\\label{Mcut}
&\chi\ge F_c+(\mathbf{T}\mathbf{y}_m-\mathbf{r})^T \mathbf{z}_{ms}^l-\mathbf{t}_e^T \bm{\lambda}_{s\xi}^l-\mathbf{t}_{ie}^T\mathbf{z}_{s\xi}^l\\\nonumber
&+(\mathbf{J}_e\bm{\xi}^l)^T\bm{\lambda}_{s\xi}^l+(\mathbf{J}_{ie}\bm{\xi}^l)^T\mathbf{z}_{s\xi}^l,\forall l=1,.,p\\\label{eMBJt}
&\mathbf{B}_e\mathbf{y}_s=\mathbf{t}_e-\mathbf{J}_e\bm{\xi}^p\\\label{ieMBJt}
&\mathbf{B}_{ie}\mathbf{y}_s\le\mathbf{t}_{ie}-\mathbf{J}_{ie}\bm{\xi}^p\\\label{Mcone}
&\Delta\mathbf{y}_{cone}+\mathbf{U}\mathbf{y}_s=-\mathbf{y}_{cone}^p\\\label{MH}
&\mathbf{y}_{coneij}^{pT}\mathbf{H}\{\mathbf{y}_{coneij}^p+2\Delta\mathbf{y}_{coneij}\}\leq0,\forall(i,j)\in\mathbf{\Omega}\cup\mathbf{\Omega}^0\\\label{Mxc}
&\chi \ge \mathbf{F}_s^T\mathbf{y}_s+F_c
\end{align}
where \eqref{Mcut} represents all optimality cuts, which are evaluated from every dual slave till the $p^{th}$ iteration of BD. In order to deal with the slow convergence issue of master sub-problem and to direct its solution towards optimality, an acceleration strategy is used to include the linearized form of all constraints in \eqref{Gineqxi} - \eqref{Gcone} with known $\bm{\xi}=\bm{\xi}^p$ in the master sub-problem. Fortunately, \eqref{Gineqxi} - \eqref{Geqcone} are already linear, however, the constraints in \eqref{Gcone} are nonlinear. Thus, by the use of Taylor's series, \eqref{Gcone} is linearized as \eqref{MH} with other constraints being \eqref{eMBJt} - \eqref{Mcone}. \eqref{Mxc} ensures that $\chi$ also acts as upper bound for the objective of the primal slave sub-problem. To further improve the computational efficiency of the master sub-problem, the number of optimization variables in \eqref{Mobj} - \eqref{Mxc} are reduced by considering the actual auxiliary variables in \eqref{Dcii} and \eqref{Dcsij}, instead of the variables (other than the auxiliary variables) in \eqref{Dd1ij} - \eqref{Dd4ij}. The master problem in \eqref{Mobj} - \eqref{Mxc} is a Mixed-Integer Linear Programming (MILP) problem and can be solved by existing MILP solver in CPLEX. The overall BD based solution methodology to obtain the robust AC TEP plans is discussed next.\vspace{-0.3cm}
\subsection{Algorithm}
\begin{enumerate}
\item Set BD's iteration count $p=1$, lower bound $LB=-\infty$, upper bound $UB=+\infty$, and choose the desired tolerance $\epsilon$.
\item Choose an initial transmission network topology, possibly a feasible topology, which may be obtained from the deterministic AC TEP \eqref{Dobj} - \eqref{DCPRlim} with nominal loads and RES generations at respective rated capacities.
\item Solve the proposed dual slave sub-problem \eqref{MSobj} - \eqref{MSuzeta} by PDIPM to find the desired dual variables and associated $\bm{\xi}^p$ for the worst-case uncertainty realization.
\item Add optimality cut in \eqref{Mcut} associated with the feasible dual slave sub-problem to the master sub-problem.\label{step4}
\item Solve the master sub-problem \eqref{Mobj} - \eqref{Mxc} by CPLEX and update $LB$ as the objective of the master sub-problem.
\item Solve the dual slave sub-problem \eqref{MSobj} - \eqref{MSuzeta} with fixed master variables and update $UB$ as the sum of the objective of the dual slave sub-problem and the latest line investment cost from the master sub-problem.
\item If $(UB-LB)/UB\geq\epsilon$, repeat from step \ref{step4}, else stop.
\end{enumerate}
The proposed approach gives a robust AC TEP plan in finite number of BD iterations, which are feasible for all realizations of the uncertain parameters, as defined by the interval uncertainty sets. This is proven by the numeric results, which are discussed next. Also, this paper assumes uncertainty in all the loads and RES generations that reach their respective limits as per their interval set. This may be conservative. Alternatively, the budget of uncertainty is a promising option \cite{jabr2013robust} that can also be adopted in the proposed approach.

\section{Numeric Results}\label{sec4}
The proposed solution methodology for robust AC TEP plans is tested on a $3$ bus, Garver $6$ bus with and without initial topology ($6$-gf) \cite{yu2010robust}, $24$ bus \cite{yu2010robust}, $46$ bus \cite{romero2002test} and $87$ bus \cite{romero2002test} systems. The data for these test systems is given in \cite{data}. The maximum allowable installations $\overline{n}_{ij}$ (including base topology) of identical circuits in each transmission corridor is three for $3$ bus, $24$ bus, and $46$ bus systems, and five for $6$ bus and $6$-gf systems. In the $87$ bus system, besides the base topology, three more identical transmission circuits are allowed in each corridor. For the AC TEP problem, the line investment cost coefficients are scaled such that $\max\{IC_{ij}\}<12,\forall (i,j)\in\bm{\Omega}$. Also, the cost coefficients of the generators are a small fraction of the scaled $IC_{ij}$ so that the planning solution is not adversely affected. Further, the curtailment costs of loads and RES generations are $100$ and $500$, respectively. The Big-M number $M$ and tolerance for BD are chosen as $10$ and $10^{-5}$, respectively. The computing platform is MATLAB (for CPLEX and PDIPM) and GAMS on an Intel Core I7-8700 CPU at 3.20 GHz and 16 GB RAM.

\subsection{Validation of zero DG of proposed dual and zero Optimality Gap (OG) of AC TEP in Section \ref{2.B}}\label{sec41}
Two aspects of the AC TEP formulation in this paper are validated (with fixed topology and no RES generations)
\begin{enumerate}
\item \emph{Case 1}: Zero DG of the proposed dual slave sub-problem \eqref{Sobj} - \eqref{Slim} in BD with respect to the primal slave sub-problem \eqref{Pobj}, \eqref{Pineqxci} for deterministic AC TEP.\label{case1}
\item \emph{Case 2}: Zero OG of convex relaxation of AC TEP \eqref{Dobj1} - \eqref{Dthetaijk} with respect to the non-convex and nonlinear AC TEP \eqref{Dobj} - \eqref{DCPRlim}.\label{case2}
\end{enumerate}

To validate \emph{Case \ref{case1}}, the dual of the primal slave sub-problem \eqref{Pobj}, \eqref{Pineqxci} for deterministic AC TEP by CDT \cite{anderson2016duality} (associated formulation is given in Section 4 of \cite{proof}) is solved in GAMS by KNITRO for the above test systems with two topologies: first, no additional lines over base topology, and second, few existing additional lines over base topology to form a feasible topology. Corresponding results are given in columns 7 - 9 of the table I in \cite{results}. Under similar conditions, the proposed dual \eqref{Sobj} - \eqref{Slim} is also solved in MATLAB by PDIPM, and the obtained results are shown in columns 10 - 12 of the table I in \cite{results}. Similar results from the associated primal sub-problem in deterministic AC TEP, i.e., \eqref{Dobj1} - \eqref{Dthetaijk}, solved in GAMS by KNITRO, are shown in columns 4 - 6 of the table I in \cite{results}. For these fixed topologies, objective values, total generation $P_g$ and load curtailment $CP_d$ are given in the table I in \cite{results}. The objective value is the sum of the total generation and the load curtailment costs. It is scaled by $8760$ to represent the total cost in \$/yr. It can be observed that, in all test systems, the proposed dual provides the same objective, total generation, load curtailments as obtained in the associated primal. We have also observed that other variables in the primal and the proposed dual match each other. This proves the zero DG/ strong duality between the proposed dual and the associated primal. However, the solution from the dual obtained by CDT ends up with significant DG. It is important to make sure that the DG is zero, else infeasibility is encountered in the proposed methodology. Subsequently, to validate \emph{Case \ref{case2}}, the nonlinear and non-convex AC TEP in \eqref{Dobj} - \eqref{DCPRlim} is solved in GAMS by KNITRO. The total system load for each test system is given in column 2 of the table I in \cite{results}. It is observed that the same solution is obtained (as shown in columns 4 - 6 of the table I in \cite{results}) from \eqref{Dobj1} - \eqref{Dthetaijk} and \eqref{Dobj} - \eqref{DCPRlim} for all test systems and considered topologies, which confirms zero OG. Obtaining zero OG solution from \eqref{Dobj1} - \eqref{Dthetaijk} for deterministic AC TEP confirms that the AC TEP plans match with the plans from \eqref{Dobj} - \eqref{DCPRlim}. Similar observation is also reported in \cite{jabr2013optimization}.

\begin{table}[htbp]\vspace{-1em}
\caption{Robust AC TEP plans for different load uncertainties}
\renewcommand{\arraystretch}{0.77}%
\renewcommand{\tabcolsep}{0.5mm}
\begin{center}
\begin{tabular}{|c|c|c|c|c|}
\hline
\multirow{2}{*}{System} & Uncertainty & Total cost & Robust AC & Total\\
& in load ($u_d$) & ($\times10^5$\$/yr) & TEP plan & $CP_d$ $(pu)$\\\hline
\multirow{7}{*}{3 bus} & 0\% & 0.0766 & $n_{2-3}$=1 & 0 \\ \cline{2-5} 
 & 5\% & 0.1191 & $n_{2-3}$=2 & 0 \\ \cline{2-5} 
 & 10\% & 0.1395 & $n_{1-3}$=$n_{2-3}$=1 & 0 \\ \cline{2-5} 
 & 15\% & 0.1821 & $n_{1-3}$=1,$n_{2-3}$=2 & 0 \\ \cline{2-5} 
 & 20\% & 0.2438 & $n_{1-3}$=$n_{2-3}$=2 & 0 \\ \cline{2-5} 
 & 25\% & 0.7242 & $n_{1-2}$=1,$n_{1-3}$=$n_{2-3}$=2 & 0.0467 \\ \cline{2-5} 
 & 30\% & 2.0382 & $n_{1-2}$=1,$n_{1-3}$=$n_{2-3}$=2 & 0.1967\\\hline
 \multirow{7}{*}{6 bus} & 0\% & 1.1167 & $n_{2-6}$=1,$n_{4-6}$=2 & 0.0142 \\ \cline{2-5} 
 & 5\% & 1.2823 & $n_{2-6}$=$n_{4-6}$=2 & 0 \\ \cline{2-5} 
 & 10\% & 1.3619 & $n_{2-6}$=$n_{4-6}$=2 & 0.0016 \\ \cline{2-5} 
 & 15\% & 1.5005 & $n_{1-5}$=1,$n_{2-6}$=$n_{4-6}$=2 & 0.0017 \\ \cline{2-5} 
 & 20\% & 1.7127 & \begin{tabular}[c]{@{}c@{}}$n_{1-5}$=$n_{3-5}$=1,\\ $n_{2-6}$=$n_{4-6}$=2\end{tabular} & 0.0026 \\ \cline{2-5} 
 & 25\% & 2.0167 & \begin{tabular}[c]{@{}c@{}}$n_{1-5}$=$n_{3-5}$=1,\\ $n_{4-6}$=2,$n_{2-6}$=3\end{tabular} & 0.0011 \\ \cline{2-5} 
 & 30\% & 2.2382 & \begin{tabular}[c]{@{}c@{}}$n_{2-3}$=1,$n_{3-5}$=2,\\ $n_{2-6}$=2,$n_{4-6}$=3\end{tabular} & 0 \\ \hline
\end{tabular}
\end{center}
\label{un_load}\vspace{-0.9cm}
\end{table}
\subsection{AC TEP with different levels of load uncertainty}\label{ACTEP_pd}
The proposed approach is first tested for the $3$ bus system to obtain robust AC TEP plans and the curtailments under different levels of load uncertainty (both real and reactive load). The $3$ bus system consists of three corridors, with one line in each corridor, and the total real power generation capacity is $3.5pu$ with total load of $(3+j2.5)pu$. Table \ref{un_load} shows the robust AC TEP plans for different levels of load uncertainty in the $3$ bus and $6$ bus systems with fixed RES generations. For the $3$ bus system, one wind turbine at bus $2$ with $75MW$ capacity is considered. In most papers of the literature, a load uncertainty in the range of $5\% - 25\%$ is usually considered \cite{zhuo2019incorporating,yu2009chance,ruiz2015robust,dehghan2017adaptive,yu2010robust}. In this paper, load uncertainties in range of $0\% - 30\%$ are considered. For a forecasted real power load $P_{di}$ at bus $i,\forall i\in\mathbf{N}_b$, the associated interval uncertainty bounds are $\xi^{max}_{di}=-\xi^{min}_{di}=u_d P_{di}/100$, where $u_d$ is the percentage of real and reactive load uncertainty. For $u_d=0\%$ in all loads of the $3$ bus system, the total line investment cost is $4000$\$/yr and generation cost is $3664.5$\$/yr without any curtailment. Further, as $u_d$ increases, line installations increase along with the generation in order to meet the load. However, with $25\%$ and $30\%$ load uncertainties, total system load is not met, as total available generation capacity is utilized. Hence, load curtailments start appearing. Consequently, the total costs are significantly high due to high curtailment cost. The same can be observed in Fig. \ref{fig1} for $3$ bus system, where the percentage increase in total cost increases as load uncertainty increases from $5\%$ to $30\%$ with fixed RES generations. For $25\%$ and $30\%$ load uncertainties, the total costs are $845.431\%$ and $2560.836\%$ with respect to the deterministic cost ($u_d=0\%$ with fixed RES generation), respectively. Thus, the AC TEP plans so obtained, not only provide the location of line installations, but also aid in locating the loads, which have to be curtailed for the normal operation of the system. Similarly, for the $6$ bus system, different levels of real and reactive load uncertainties are considered, and the corresponding results are shown in Table \ref{un_load}. In this system also, load curtailments start appearing in few cases of load uncertainties, but not due to full utilization of total generation capacity. In fact, the curtailments appear as the associated cost is comparatively less than the cheapest line installation. \emph{Results for the $6$-gf, $24$ bus, $46$ bus and $87$ bus systems under different levels of load uncertainty, and with fixed RES generation, can be seen in table II and table III of \cite{results}}. Also, Fig. \ref{fig1} presents the percentage increase in the total cost for various case studies with respect to the deterministic case ($u_d=0\%$ with fixed RES generation) for the above test systems. It can be observed in Fig. \ref{fig1} that, for $24$ bus system with $30\%$ load uncertainty, the percentage increase in the total cost with respect to the deterministic case is very high with a value of $30339.707\%$. Further, for the $6$ bus and $6$-gf systems, three $100MW$ wind turbines at each bus $1$, $3$ and $5$ are considered. Similarly, for the $24$ bus system, three $300MW$ wind turbines at each bus $6$, $14$, $22$ are assumed to be connected. For the $46$ bus system, $500MW$, $200MW$ and $500MW$ wind turbines at each bus $16$, $24$ and $42$ are considered. For the $87$ bus system, two $50MW$ wind turbines at each bus $25$, $69$ and four $100MW$ wind turbines at each bus $2$, $35$, $44$, $50$ are considered. Also, the obtained AC TEP plans for all test systems with different load uncertainties are $100\%$ robust for all realizations of the load uncertainties (details of robustness check in Section \ref{234324}).
\vspace{-0.5em}
\subsection{AC TEP with load and RES uncertainty}
\begin{figure}[htbp]\vspace{-0.5em}
\centering
\includegraphics[width=0.5\textwidth]{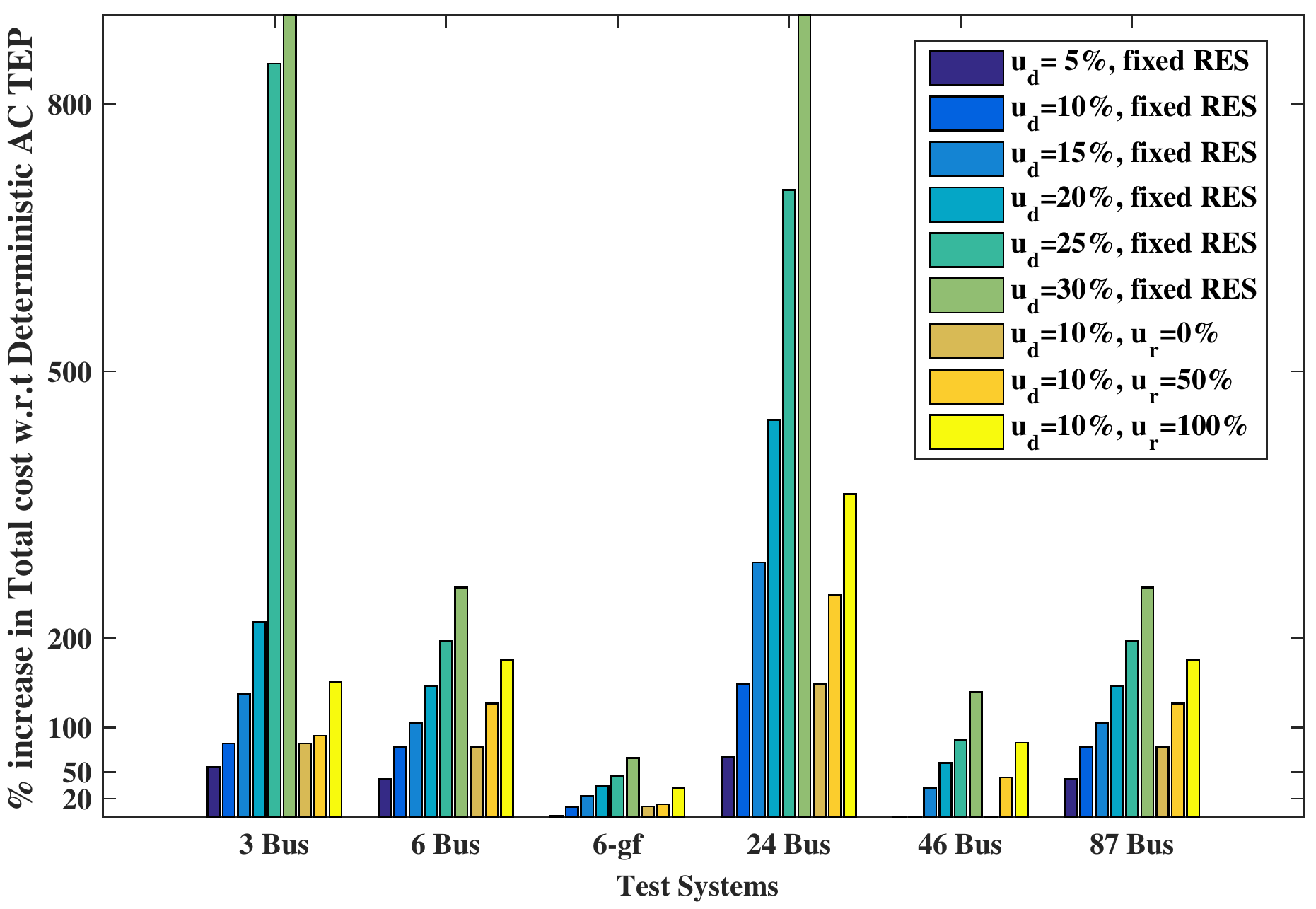}
\caption{\% increase in cost of AC TEP under different injection uncertainties}
\label{fig1}\vspace{-0.7cm}
\end{figure}

The effect of RES generation uncertainty, along with load uncertainty, is an important and practical study in planning problems. Hence, robust AC TEP plans for all test systems with both loads and RES generations uncertainties are obtained by the proposed approach. The RES generation data considered is same as given in section \ref{ACTEP_pd}. For a forecasted real power RES generation $P_{ri}$ at bus $i,\forall i\in\mathbf{N}_b$ and percentage uncertainty in RES generations $u_r$, the associated RES generation interval uncertainty bounds are defined as $\xi^{max}_{ri}=0$ and $\xi^{min}_{ri}=-u_r P_{ri}/100$. This is because, unlike the load uncertainty, a RES generation generally does not exceed its rated capacity, even with RES generation uncertainty. \emph{Robust AC TEP plans for all the test systems with $10\%$ load uncertainty in all loads and three levels of RES generation uncertainties ($0\%$, $50\%$, $100\%$) with the CPU run times are given in table IV of \cite{results}}. High levels of RES generation uncertainty are considered to represent practical scenario, where the RES generation may be completely zero or may reach half of its capacity. It is observed that, for the given load uncertainty and with increase in RES generation uncertainty, new installations and total generation also increase. For instance, in the $3$ bus system, one additional line in each of the corridors $1-3$ and $2-3$ is required, when $10$\% load uncertainty with fixed RES generation ($u_r=0\%$) is considered. The associated line cost is $10,000$\$/yr and generation cost is $3952.5$\$/yr. As RES generation uncertainty increases, the proposed approach provides robust AC TEP plan so that the worst-case protection within the uncertainty range is guaranteed. Hence, with increase in the RES uncertainty, the worst-case scenario depicts a decrease in RES generation, which however gets compensated by the conventional generators while catering to the loads. For these cases, the load and RES curtailments in all test systems are zero. On similar lines, robust AC TEP plans for other test systems have also been obtained and are given in table IV of \cite{results}. The percentage increase in total cost with $u_d=10\%$ and different levels of RES generation uncertainties with respect to deterministic cost ($u_d=0\%$ with fixed RES generation) for all test systems is shown in Fig. \ref{fig1}.

\subsection{Robustness of AC TEP plans}\label{234324}
The robustness of the obtained AC TEP plans, for all the case studies corresponding to each test system in Table \ref{un_load} and the tables II-IV in \cite{results}, is verified by MCS. Random samples of the associated uncertain parameters are generated within their respective interval uncertainty sets and actual nonlinear and non-convex ACOPF is solved by PDIPM \cite{torres1998interior} for each sample by freezing the obtained robust AC TEP plan for each case. This ACOPF includes load and RES generation curtailment options. In this paper, $16,600$ samples \cite{jabr2013robust} are uniformly generated within the uncertainty set for each case, and each sample is checked for the convergence of ACOPF and also for the constraints satisfaction. It is observed that the nonlinear and non-convex ACOPF always converges for each sample and for each test case. Thus, it is concluded that the AC TEP plans from the proposed bi-level BD based approach are $100\%$ robust for all the considered scenarios, as defined by the injection uncertainties. Additionally, the proposed AC TEP formulation always ensures feasibility, as it includes both load and RES generation curtailment options. Hence, any issues related to the convex relaxation in section \ref{2.B} are inherently taken care and are automatically adjusted via the actual nonlinear and non-convex ACOPF in MCS based verification of the robust AC TEP plans.\vspace{-0.5em}

\section{Conclusion}\label{sec5}
\emph{In this paper, a first bi-level BD based RO solution approach is proposed for AC TEP with load and RES generation uncertainties}. A well established convex relaxation of AC TEP is adopted in the proposed approach. \emph{The novelty of the proposed approach are the use of: 1) a unique and novel dual in second level of BD that ensures strong duality, overcomes the drawback of the dual by CDT, and is solved by PDIPM, 2) novel linear constraints in the proposed dual for the worst-case realization of uncertainties that avoid resorting to MILP solvers, and 3) additional constraints, with the use of Taylor's series, in the first level of BD to speed up the convergence of the overall process}. The approach is tested on various test systems and it is observed that the total cost of robust AC TEP with load uncertainty increases with higher level of uncertainty. Further, with RES penetration and increase in its uncertainty level, the total cost increases. Also, the obtained AC TEP plans are $100\%$ robust, as evidenced from the MCS based robustness check. Thus, the robust AC TEP plan from the proposed approach proves to be feasible for all the uncertain realizations, within the uncertainty set. Additionally, the robust plans are accompanied with the minimization of load and RES generation curtailments, thereby ensuring high priority to consumers and encouraging green power utilization, respectively. The over-conservative nature of the obtained plans can be reduced by easily including budget of uncertainty constraints in the proposed approach. It may be interesting to explore the proposed method for multi-stage and coordinated transmission and reactive power planning.\vspace{-0.3cm}
\bibliography{IEEEabrv,Dualref}
\end{document}